\documentclass{article} \usepackage{epsfig}

 \newcommand{\sect}[1]{\S\ref{sect.#1}}  
 \newcommand{\eq}[1]{Eq.~(\ref{eq.#1})}
 
 \newcommand{\fig}[1]{Fig.~\ref{fig.#1}}

 \newcommand{\sectlabel}[1]{\label{sect.#1}}
 \newcommand{\eqlabel}[1]{\label{eq.#1}}

\newcommand{\figdef}[3]{
\begin{figure}[!htb]
 \centering\leavevmode#2%
 \caption{\small #3}
 \label{fig.#1}
\end{figure}                 }

\newcommand{\expect}[1]{
\left\langle #1 \right\rangle
}

\title{\vspace{-0.7in}Quantum Smart Matter}

\author{\vspace{0.1in}Tad Hogg and J. Geoffrey Chase\\ Xerox Palo Alto
Research Center \\ 3333 Coyote Hill Road \\ Palo Alto, CA 94304,
U.S.A. \\ hogg@parc.xerox.com, chase@parc.xerox.com}

\begin{document}

\maketitle

\vspace{-0.1in}
\begin{abstract}
  \vspace{0.05in}
  \noindent The development of small-scale sensors and actuators
  enables the construction of ``smart matter'' in which physical
  properties of materials are controlled in a distributed manner. In
  this paper, we describe how quantum computers could provide an
  additional capability, programmable control over some quantum
  behaviors of such materials. This emphasizes the need for spatial
  coherence, in contrast to the more commonly discussed issue of
  temporal coherence for quantum computing. We also discuss some
  possible applications and engineering issues involved in exploiting
  this possibility.
\end{abstract}

{\em A condensed version of this paper will appear in the PhysComp96
conference proceedings.}

\section{Introduction}

Distributed control, using many sensors, computers and actuators, can
improve the performance of systems at many scales.  Examples include
controlling traffic flow in cities~\cite{mahmassani87}, regulating
office environments~\cite{huberman95b}, active strengthening of
structural materials~\cite{berlin94}, structural vibration
control~\cite{how92}, reducing fluid turbulence~\cite{berlin95} and
adjusting optical responses~\cite{mcmanamon96}.  The continuing
development of micrometer-scale machines~\cite{bryzek94} and proposals
for even smaller devices~\cite{drexler92} constructed with atomically
precise manipulations~\cite{bell96,eigler90,jung96,shen95} offer
further possibilities for designing materials whose properties can be
modified under program control, giving rise to so-called ``smart
matter''~\cite{kersey96}.

Smart matter is a material that locally adjusts its response to
external inputs through programmed control. Such control is enabled by
embedding sensing, computation and actuation ability within the
material. Specifically, control programs are designed to use
measurements of the system response to compute appropriate control
inputs to the system, such as forces or electric fields, which are
then imposed on the system by the actuators. This operation is known
as feedback control because measurements of the system response are
fed back to the controller for use in determining the control inputs.

To date, proposals for smart matter focus on controlling classical
behaviors of materials~\cite{berlin94,kersey96}. However, small,
precisely constructed devices can also exploit quantum
behaviors~\cite{alivisatos96,debreczeny96,nguyen96,rosencher96}. Thus,
an interesting open question is the extent to which the demonstrated
abilities to modify quantum behaviors, together with distributed
computation, can provide a much finer level of control of the
properties of a material. This leads to {\em quantum smart matter},
which consists of actuators, sensors and computers integrated to
operate on quantum behaviors.

In this context, classical control methods and computers have a
limited role due to their use of a measurement process which
necessarily disrupts the quantum behavior.  Instead, control of the
quantum behavior of materials is a possible application for quantum
computers~\cite{benioff82,bernstein92,deutsch85,deutsch89,divincenzo95,feynman86,lloyd93}. Although
there has been some work on distributed, parallel quantum
computers~\cite{margolus90}, the use of quantum computers for
controlling materials contrasts with most studies of such computers,
which focus on purely computational questions such as whether they can
compute classically intractable functions.  Quantum computers are
distinguished by their ability to operate simultaneously on
superpositions of many classical states (``quantum parallelism''), and
their restriction to unitary linear operations on such superpositions
which can be used to produce interference among different
computational paths. In particular, this restricts the programs to be
reversible, and hence requires development of reversible
devices~\cite{merkle92}.

In this paper, we discuss coupling the programability of quantum
computers to properties of materials, to create quantum smart
matter. Both classical and quantum smart matter share the basic idea
of using a large number of integrated sensors, computers and
actuators. They differ in that using quantum computers avoids the need
to perform measurements on the quantum system. In the remainder of
this paper, we first describe some of the control options for smart
matter, then present an idealized example, and discuss a number of
possible applications.

\section{Types of Control}

\subsection{Global and Distributed Controls}\sectlabel{controls}

A large majority of the control applications implemented today use
global controls~\cite{boyd91,doyle92,slotine91,vidyasagar93}. These
controllers employ a single centralized controller that receives
measurements of the system's state and delivers control inputs. Their
popularity stems from their conceptual simplicity: the control program
deals directly with the desired overall properties of the system and
need not coordinate its activities with other controllers. More
formally, existing theoretical tools provide a basis for establishing
provable performance bounds and the optimal use of control resources.

Global controllers have serious drawbacks in the context of smart
matter. First, manufacturing defects and variations in the environment
make it difficult to accurately model the exact dynamic behavior of
the system.  Second, coordinating the activities of all the actuators
in real time becomes an intractable designing and programming task as
the number of active elements (sensors and actuators) increases. There
can also be communication bottlenecks from the need to provide all the
system measurements to the central controller in a timely
manner. Finally, the failure of the single central controller
completely eliminates all control of the system.

These difficulties motivate the use of distributed, or decentralized,
control mechanisms. These control methods consist of a combination of
many controllers, each designed and operated with limited knowledge of
the complete system. This approach can allow control to be applied to
more complex systems, including distributed
computation~\cite{huberman88}. While global performance cannot
necessarily be guaranteed as with global controllers, decentralized
controllers can, in practice, be remarkably robust to the failure of
individual active elements, and are found in a variety of systems such
as biological ecosystems, market economies and the scientific
community. Some applications of distributed control include regulating
office environments~\cite{huberman95b}, traffic
flow~\cite{mahmassani87}, and, in the context of smart matter,
structural vibrations~\cite{how92}.

\subsection{How Control Can Make Smart Matter}

Materials with desirable properties can be created in a number of
ways. For most materials in use today, the properties are built in
through a suitable choice of component materials and fabrication
method (e.g., plastics and metal alloys). This technique is very
robust when suitable materials can be found, but limited by properties
of natural materials the available fabrication technologies. In
effect, this procedure designs the system so additional control is not
needed, i.e., the uncontrolled behavior of the system has the desired
properties already. Unfortunately, once fabricated it is difficult to
change the material properties. This is especially true of changes
that should take place only at specific locations and occur rapidly in
response to some environmental change.

One way to change the properties of materials in a controlled manner
is through the use of external fields applied to the whole
system. Provided the relevant physical properties change in response
to this field, changing the field provides a global control of the
material. Examples include piezoelectric crystals where electric
fields modify mechanical properties and the use of lasers to modify
chemical reactions of large groups of
molecules~\cite{chen95,dahleh96,gross93,zhu95}. If the system can be
accurately modelled these external fields can be designed a
priori. However, designing effective global control for large,
dynamic, heterogeneous systems is intractable due to the scale and
difficulty of modelling their quantum behavior. Alternatively, if many
repeated experiments are feasible, the controls can be adapted to the
system by incremental changes that improve performance based on
measurements of the system response.

Another approach is to apply the required fields locally through
embedded actuators, but still without any sensors. This alternative
can handle spatial variations in the material that are known in
advance, or provide a match to a fixed system through overall
adaptation after many trials. This alternative still does not
dynamically adjust to variations resulting from imperfections in the
system.

The above control methods work without any feedback, either by having
good knowledge of the system behavior so the control force can be
suitably designed, or through an adaptive process where different
controls can be applied to many copies of the system to determine
which method is best. When these conditions do not hold, these control
methods are not effective.

Smart matter, where sensors and actuators are integrated in large
numbers throughout the material, leads to an an alternate control
method: the ability to sense and act independently on a local scale is
employed to create desirable global behavior.  This approach allows
the control force to respond dynamically to unanticipated changes in
the system or compensate for an inaccurate model of the dynamics at a
very local scale.  In effect, this allows the adaptation to take place
while the system operates and in response to local variations, in
contrast to a global adaptation of controls without local sensors
where adjustments are based on the average behavior of many trials or
copies of the system. A good example of the need for dynamic control
is the behavior of vortices near a surface moving through a turbulent
fluid. Here local sensors can allow response to individual vortices,
whose location and occurrence are not readily predicted.
 
The most extreme case of smart matter is when the computation needed
for the control is fast compared to any relevant changes in the
physical configuration of the material. This allows for a decoupling
of the slow physical degrees of freedom, which we denote by $P$, from
the rapid computational degrees of freedom, denoted by $C$, in the
same way that molecular or solid-state dynamics can often be
approximated by considering separately the behavior of the electrons
and the atomic nuclei. For example, this could be achieved by using
light particles for the computation while heavy ones determine the
relevant physical response.  Viewed another way, within a given
implementation, this requirement also limits the number of
computational steps that the control program can perform to determine
its result, thus, defining the maximum acceptable latency of the
control system.

Finally, the distinction between a variety of individually fabricated
materials and smart matter, where properties can be changed under
program control, is somewhat analogous to the distinction between
customized electronic circuits for specific tasks and the use of
general microprocessors. In the former cases, the customized material
or circuit have a fixed set of properties, and can be well-matched to
specific applications whose requirements do not change rapidly. For
the latter, the programmability of smart matter or microprocessors,
allows for a wider range of applications and flexible response to
changes.

\subsection{Controls for Quantum Smart Matter}

Controlling quantum behaviors elegantly extends the capabilities of
smart materials, since the active elements can operate with the full
quantum state of the material. Realizing this possibility requires
translating classically based control methods to quantum
systems. These methods include controlled behavior based on either
feedback or
modelling~\cite{boyd91,doyle92,slotine91,vidyasagar93}. The difficulty
of applying these control techniques will depend on the way the system
is constructed.

At one extreme, precise construction~\cite{drexler92} simplifies the
control problem by allowing accurate modelling of the environment of
each device and individually tailored programs, but imposes severe
difficulties for fabrication.  On the other hand, more readily
manufactured devices will exist in a statistically variable
environment, making the control design more difficult. It is this
latter case that we mainly focus on here, as it raises a number of
engineering control issues where sophisticated controls can compensate
for current inability to precisely fabricate materials.

One consequence of employing localized control is that creating
macroscopic effects with microscopic controllers will require a large
number of controllers. An immediate consequence is the requirement
that these controllers be relatively homogeneous in design and
function to simplify their design and construction. Furthermore, each
controller will be required to act either autonomously, or in concert
with only a few others, since the design of complex interactions among
so many controllers using a global model and algorithm is intractable.
Thus, quantum smart matter will be based on controllers, designed with
local knowledge and behavior, which are homogeneous and act
autonomously to achieve a desired macroscopic effect.

A second consequence is that global simulation and performance
predictions will be statistical in nature due to the inability to
precisely specify each controller's detailed environment, or perform
simulation and optimization for so many degrees of freedom. In fact,
the precise location of the devices would be described according to a
probability distribution rather than known a priori, as assumed by
standard control methods. This fact will require different types of
systems analysis than are typically employed with classical systems.

In a control context, the forces acting on the physical degrees of
freedom of a material must also depend on the computational ones.
This observation is the analog of actuators in classically defined
smart matter where results of a computation can change the forces
acting on the physical system. Thus, the potential acting on $P$ must
be a function $V(P,C)$.  Within the range of variation of this
function, the control program can adjust $C$, based on the value of
$P$, to produce an effective physical potential defined by

\begin{equation}
V_{\rm eff}(P) = V(P,C(P))
\eqlabel{potential}
\end{equation}

\noindent Because a quantum computer can perform this operation on
quantum superpositions, it produces a system whose relevant physical
behavior is governed by $V_{\rm eff}(P)$, a controlled potential for
the quantum system.

This discussion illustrates the difference between quantum smart
matter, controlled by quantum computers, and smart matter whose
control is determined by a classical computer. Since quantum computers
operate on superpositions they are essentially applying all possible
control actions weighted by the wave function values. Hence, quantum
smart matter does not need to measure quantum states for feedback By
contrast, classical computers can't give feedback control since they
would require a measurement of the state, hence collapsing the wave
function.

\eq{potential} also illustrates a similarity between quantum and
classical control of smart matter: both types of control modify the
potential governing the system's dynamics.  Quantum smart matter does
this for quantum systems, without collapsing the wave function, while
classical feedback modifies classical dynamic system properties.

Another important control issue is that of stability: whether the
control achieves the desired behavior while maintaining the state of
the system near some equibribum configuration. Quantitative criteria
for stability are given by results from control
theory~\cite{vidyasagar93}. For the case where the control force does
not depend explicitly on time, stability amounts to a bound on the
total energy of the system. More specifically, for reversible,
non-dissipative, quantum systems stability implies that the total
energy of the system is constant in the absence of external inputs.

These results can apply much more generally, e.g., when the control
force has an explicit time dependence, through the use of Lyapunov
functions~\cite{vidyasagar93}. A Lyapunov function $\Lambda(P,t)$ is a
function of the system state, including the time dependence introduced
by adjustments made by the control program.  It may also have an
explicit time dependence.  If $\Lambda$ is a convex function of $P$
within some region including the initial state of the system,
$\Lambda$ has a non-positive total derivative with respect to time,
and $\Lambda(P_0)=0$ at some point $P_0$ within that region, then the
entire system will be stable with control~\cite{vidyasagar93}. In
essence this criterion determines when control feedback could
positively amplify small perturbations in the system state, resulting
in unstable behavior.

This result suggests a direct link between the potentials, $V(P,C)$,
and classical control theory in which there are several synthesis
methods that guarantee stability and performance of the controlled
system.  Such synthesis techniques would be employed to design the
control program computing values for $C$ such that it is reversible
and satisfies the stability conditions for $V_{\rm eff}$.  Thus,
controlled stability could be guaranteed for individual systems.
However, stability of each individual system does {\it not} guarantee
stability of the entire macroscopic system, unless each system is
entirely autonomous. As an example, instability could arise when a
small change in one part of the system gives rise to larger changes in
other parts. Furthermore, the theory for control stability developed
for classical systems will need to be extended to account for quantum
systems where the long time behavior can be very different from the
classical counterpart~\cite{hogg83}.

This discussion indicates that while design and synthesis of
individual controllers may be put into a form that guarantees some
measure of local stability and performance, global stability is not as
straightforward a problem.  Such global stability measures would
likely differ from those of standard control theory in being a
statistical expectation of stability, rather than a specific proof.
Stability in this sense is particularly desirable under the assumption
that the undesirable interaction of enough elements could reduce the
expectation of stability for the entire system, and introduce
undesirable behavior. Hence, the interactions present at the
microscopic scale to which control is being applied would have to be
determined, estimated, or implicitly considered because the
application of localized controllers to achieve global results is
inherently affected by the strength of (potentially) unmodelled
interactions.

\section{Programmable Quantum Behavior}

\figdef{potentials}{
\epsfysize=1.5in
\hspace*{\fill}
\epsffile{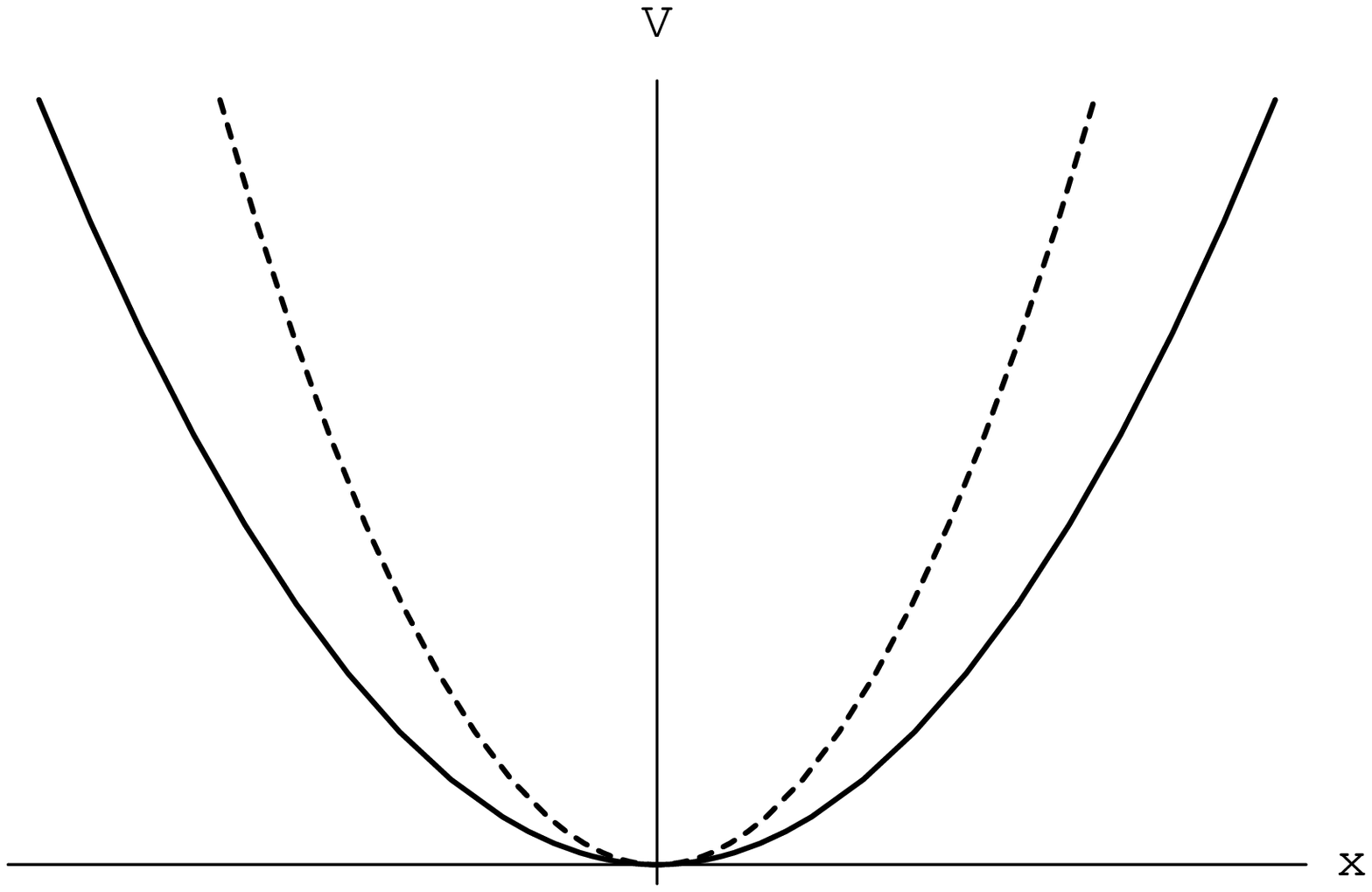}
\hspace*{\fill}
}{Example potentials $V(x,C)$ for use with quantum smart matter. 
Two cases, $V(x,0)$ and $V(x,1)$, are shown by the solid and dashed curves 
respectively, corresponding to a system with two computational states.}

Consider a one-dimensional system with a single extra
binary degree of freedom. In this case the physical degree of freedom
is the position, i.e., $P=x$, and the computational degree of freedom
$C$ is 0 or 1.  A simple potential would be to have two distinct
functional forms $V(x,0)$ and $V(x,1)$, as shown in \fig{potentials}.
Suppose the system is initially prepared in the state $|\psi \rangle =
\sum \psi(x) |x,0\rangle$ and then acted on by a quantum computer
whose program sets $C$ to 0 or 1 depending on whether $x \le 0$ or $x
> 0$, respectively. This gives

\begin{equation}
	|\psi \rangle = \sum_{x \leq 0} \psi(x) |x,0\rangle + \sum_{x > 0} \psi(x) |x,1\rangle
\end{equation}

As this state evolves, amplitudes from positive and negative values of
$x$ become mixed and the control computation acts to readjust
$C$. Provided this computation is fast compared to the physical
evolution, this gives in effect the behavior governed by the
programmed potential. Other effective potentials could be constructed
with different programs to compute $C$ from $P$.

In this example, the result is an asymmetric potential as has been
studied in the context of second harmonic generation in
optics~\cite{rosencher96}. This illustrates the trade-off between
smart materials and direct construction of materials with the desired
effective potential. On the one hand, the properties of smart
materials can be readily modified simply by changing the program in
the control computers. On the other hand, direct implementation in
materials allows for a faster response but becomes more difficult as
more complex or time-variable potentials are considered.

As a final comment on this example, note that it made use of quantum
parallelism but no use of interference. The latter capability of
quantum computers is crucial for their possible improvement on
classically intractable problems, and provides additional
possibilities for designing behavior of smart matter. For instance,
the use of destructive interference could be used to cancel the
amplitudes of certain undesirable behaviors, a feature that is not
possible with classical computations, even if they are probabilistic.

So far, we have described the behavior of a single quantum
controller. For use in smart matter, we would have a system consisting
of a large number of such devices to allow distributed control over
specific quantum behaviors of the material. Results could include
programs that provide different behaviors at different spatial
locations, as well as changing with time. Furthermore, the control
computation could make use of some of the computational states of its
neighbors, providing a way to build correlations among different
regions of the material. As an example of how simple programmed
couplings can lead to more complex potentials suppose we start with
two independent one-dimensional systems whose individual potentials
take the form $V(x,C)$ shown in~\fig{potentials}. The overall
system potential is then

\begin{equation}
V(x_{1},C_{1},x_{2},C_{2}) = V(x_{1},C_{1}) + V(x_{2},C_{2})
\end{equation}

\noindent
We can couple the behaviors together with a control program that, for
instance, sets $C_{1}$ to 0 or 1 depending on whether the {\em other}
system state satisfies $x_{2} \le 0$ or $x_{2} > 0$, respectively, and
vice versa for $C_{2}$.

In summary, quantum smart matter relies on potentials that can be
adjusted by their dependence on degrees of freedom that can be rapidly
changed under program control. To maintain superpositions, the control
computations must rely on quantum computers. Finally, if the promise
of quantum computing is realized, this capability could be used to
create complex, varying potentials governing the behavior of matter.

\section{Examples}

If implemented, quantum smart matter provides the {\em capability} for
using local control to produce desired behaviors. However, beyond the
difficulty of fabricating such systems, there remains the challenge of
{\em designing} suitable control algorithms.

\subsection{Controlling a Single Harmonic Oscillator}

To illustrate possible control methods, consider the behavior of a
one-dimensional harmonic oscillator subjected to an additional control
force so the effective potential is

\begin{equation}
V(x) = \frac{1}{2} \omega^2 x^2 + V_c(x,t)
\end{equation}

\noindent
For simplicity we restrict the control potential to be of the form
$V_c(x,t) = \frac{1}{2} k x^2 - f(t) x$, with the corresponding
control force given by $F_c(x,t)=-k x + f(t)$. The first term in this
control potential is a time-independent force proportional to $x$,
which just changes the basic frequency of the system to be $\Omega =
\sqrt{\omega^2 + k}$. The second term gives a time-dependent control
force that acts equally on the whole system, i.e., has no $x$
dependence.

With these choices, the behavior of the wave function is readily
determined~\cite{feynman65}, and is particularly simple for gaussian
wave packets, i.e., wave functions of the form

\begin{equation}
\psi(x) =\frac{1}{\sqrt{\sqrt{2 \pi} \sigma}} e^{-(x-p)^2/(4 \sigma^2)}
\end{equation}

\noindent
where $p=\expect{x}$ denotes the position of the center of the packet
and $\sigma=\sqrt{\expect{x^2}-\expect{x}^2}$ characterizes its
spread.  As the system evolves from this initial state, the wave
function continues to be described as a gaussian packet whose
position, width and phase vary with time. In particular, the position
of the center of the packet at time $T$ is given by

\begin{equation}
\eqlabel{position}
p(T) = p(0) \cos(\Omega T) + \frac{1}{\Omega} \int_0^T{ f(t) \sin(\Omega(T-t)) dt}
\end{equation}

In this context, designing a control amounts to finding values of $k$
and $f(t)$ to achieve desired behaviors. The portion of the control
force that does not depend on $x$, i.e., $f(t)$, can be delivered
either from an external global source or through the computations of
the local controller.  The forcing term $f(t)$ can be determined this
way because it does not involve any knowledge of the system state and
hence does not require any sensor values. However, providing a force
that does depend on $x$, in this case a modification of the
oscillation frequency through the value of $k$, requires the applied
force to depend on the system state.  For a classical system this
result would require the controller to measure the state for use in
its control computations. Employing quantum computers for control of
quantum systems, however, means that the controller acts on all
possible states of the system, through quantum parallelism.

Often many choices of the control force will achieve the same
objective. In these cases, additional criteria can be added to the
design. A common additional criterion is to pick from among the
feasible controls, i.e., those that produce the desired behavior, the
one that minimizes some measure of the applied control force, e.g.,

\begin{equation}\eqlabel{criterion}
\int_0^T \expect{F_c(x,t)^2} dt
\end{equation}

\noindent
This constraint acts to reduce the control gain required, in the
control design process, and therefore the actuation authority
required.  From a practical standpoint small gain controllers are
desirable since any system noise encountered undergoes minimum
amplification in the feedback control process.

For example, suppose we want the system's position, or more precisely
the expected value of the position, to be at a desired value at a
given time, i.e., we want $p(T)=\hat{p}$ at a particular time
$T$. This task is accomplished without sensors assuming accurate
information about the system parameters and the dynamics can be
integrated, as given in this case by \eq{position}. Under these
conditions, sensors are not needed to determine how the system will
behave and we can get the optimal behavior. With the explicit result
of \eq{position} and knowledge of the system parameters, in this case
the frequency $\omega$ and the initial position $p(0)$, we can obtain
the desired control by choosing $k$ and $f(t)$ such that
$p(T)=\hat{p}$. The choice that minimizes \eq{criterion} can be
determined with standard variational techniques~\cite{arthurs75} to be
$k=0$ and

\begin{equation}\eqlabel{force}
f(t) = \frac{4 \omega^2 \left(\hat{p} - p(0) \cos(\omega T)\right)}{2 \omega T - \sin(2 \omega T)} \sin(\omega (T-t))
\end{equation}

\noindent
An example of the resulting behavior for $p$ is shown in \fig{position}.

\figdef{position}{
\epsfysize=1.5in
\hspace*{\fill}
\epsffile{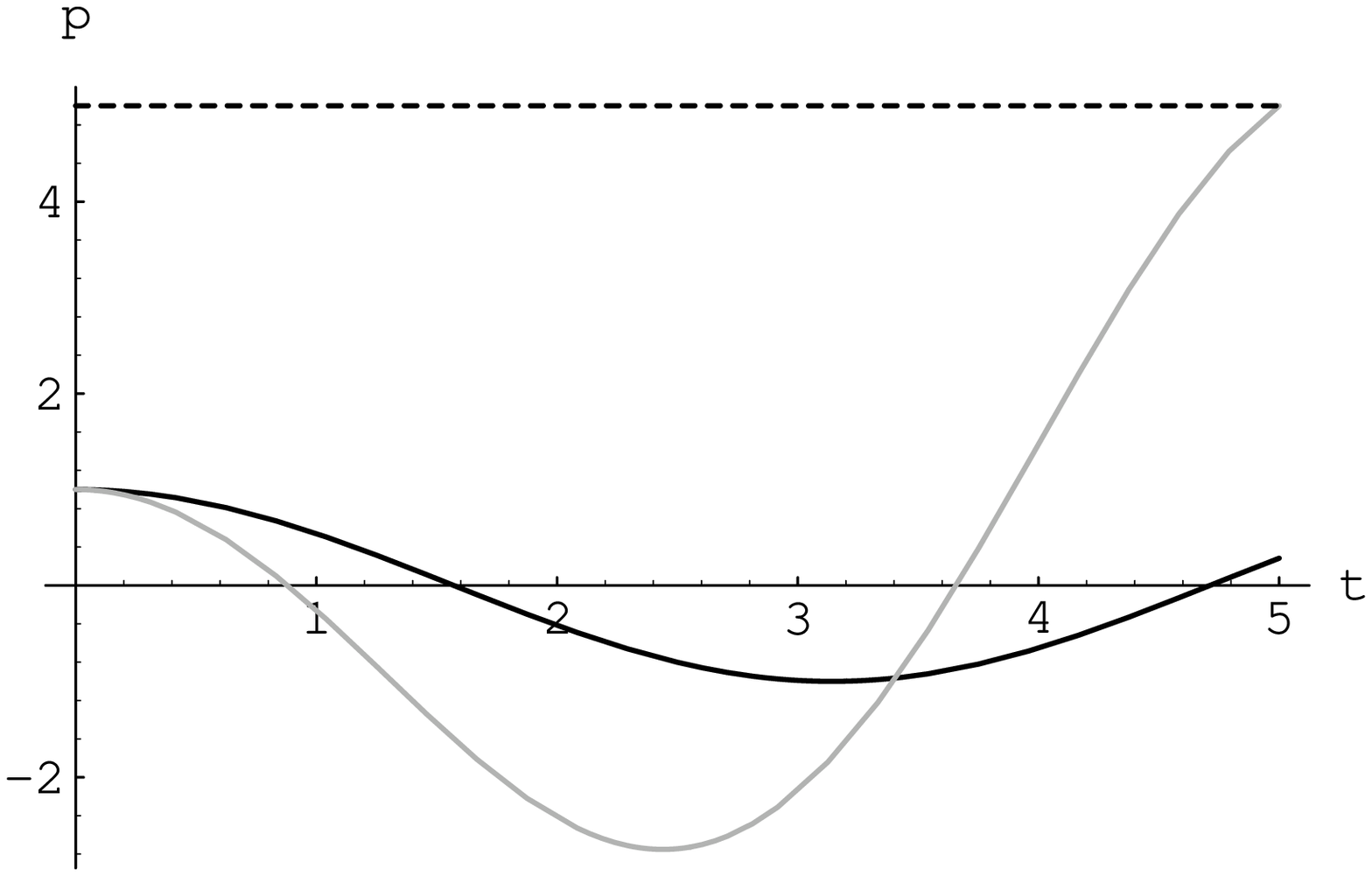}
\hspace*{\fill}
}{Expected position of a gaussian wave packet with (gray curve) and
without (black curve) control. Here the parameters are $\omega=1$,
$\hat{p}=5$, $p(0)=1$ and $T=5$. The dashed line shows the desired
final position of the packet $\hat{p}=5$.}

This example shows how a control without feedback can correctly
produce desired behaviors provided the system is accurately modeled,
and it can be solved to determine the dynamical behavior. More
realistically, we may have information on the nominal characteristics
of the material, but various imperfections in the fabrication process
or environment will cause the actual system to vary from the ideal
case. In addition, anharmonicities in the potential will make it very
difficult to integrate the dynamics even if the exact system
parameters were known. Thus, as described in \sect{controls}, this
control method will not work as well when applied to more realistic
systems.

Feedback control using sensors can address these problems to some
extent. For instance, suppose we are attempting to control to a
specific path $p(t)$, such as the one shown in \fig{position}, to
reached a desired value $p(T)=\hat{p}$, by using a force $f_{\rm
ideal}(t)$ given in \eq{force}. If the system behavior were perfectly
known, the actual system would follow this path, according to
\eq{position}. Imperfections in the model or its evaluation will cause
the system to deviate from this path. One way to address this is to
add a feedback control force of the form $-\alpha(x-p(t))$. For the
symmetric wave packets treated here, this additional force would have
zero expected value if the system matched the modeled behavior. The
overall control potential becomes

\begin{equation}
V_c(x,t) = \frac{1}{2} \alpha x^2 - (\alpha p(t) + f_{\rm ideal}(t)) x 
\end{equation}

\noindent Even if the system dynamics is only approximately known, a
large value for $\alpha$ will keep the system fairly close to the
ideal path. On the other hand, this also means the controllers are
using stronger forces, hence increasing the value of
\eq{criterion}. This illustrates a general trade-off that feedback
control provides: better performance when the system behavior is not
known precisely, but at the expense of larger control forces. An
example is shown in \fig{feedback}.

\figdef{feedback}{
\epsfysize=1.5in
\hspace*{\fill}
\epsffile{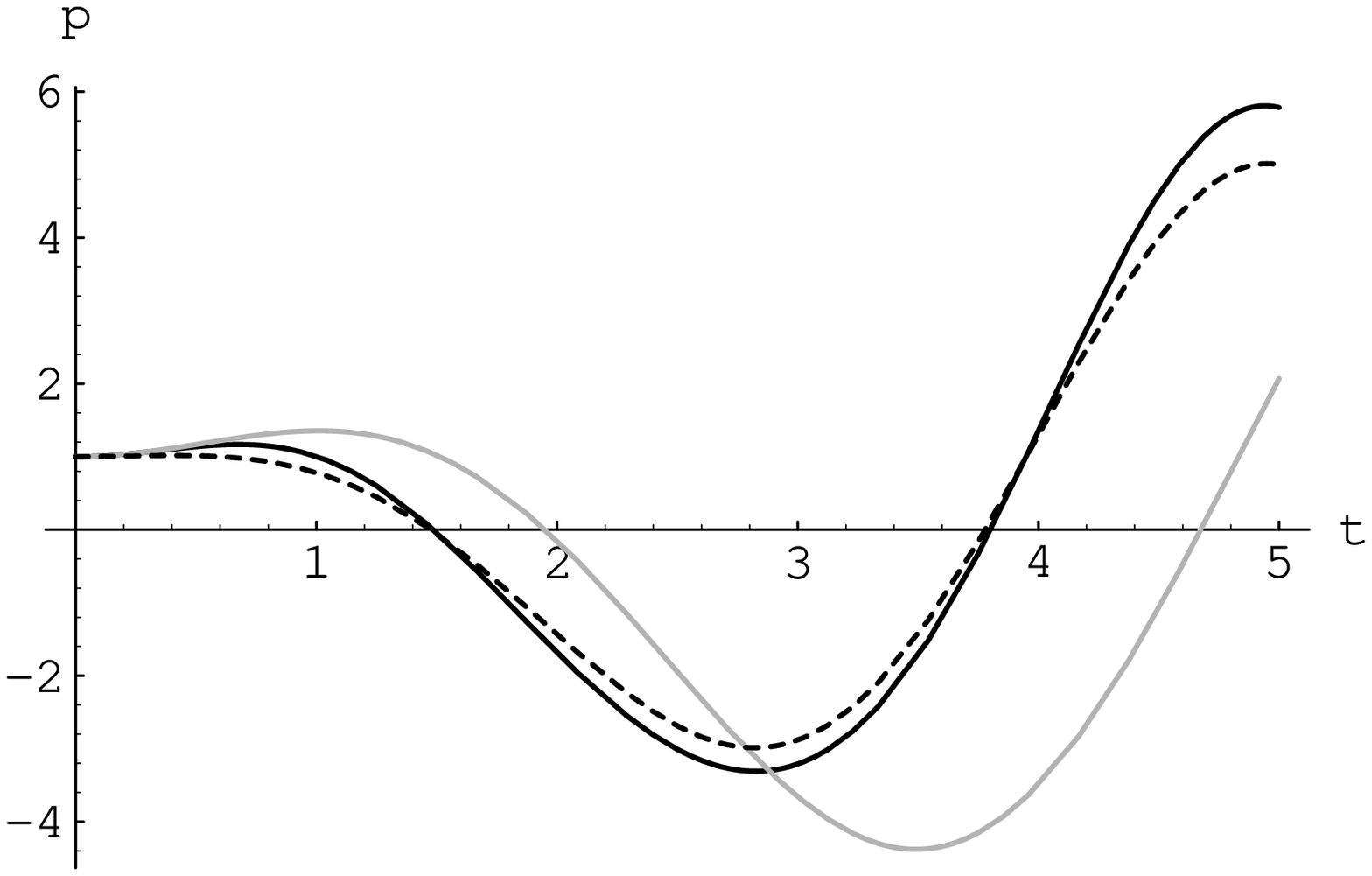}
\hspace*{\fill}
}{Using feedback to compensate for an imperfect system model. Expected
position of a gaussian wave packet with (black curve) and without
(gray curve) the addition of feedback control. The dashed curve is the
ideal path that would have been followed without feedback if the
system exactly matched the assumed model.  The parameters are
$\omega=1$, $\hat{p}=5$, $p(0)=1$ and $T=5$. The control force is
determined from the incorrect assumption that $\omega=1.5$ and
feedback uses $\alpha=10$.}

\subsection{Other Control Methods}

In addition to providing more robust compensation in the presence of
imperfect system models, feedback control (i.e., forces that depend on
the value of $x$) can be used to modify the shape of the potential
which governs dynamic behavior.  One example of using this ability is
to manipulate the width of the wave packet, something not possible for
a constant applied force. Furthermore, by changing the effective
spring constant of the system, the control can change the energy level
spacing, and thus the frequencies with which the system will
resonate. Near resonance, a small change in $\Omega$ can produce a
large change in the system response. This result represents a case
where small control forces can have a relatively large
effect. Specifically, near resonance, the size of the system response
to an external force of frequency $w$ is proportional to $1/(\Omega -
w)$. Thus, small changes in the value of $k$ in the controller design,
and the consequent small changes in $\Omega$, can lead to large
changes in the system response. This change could in turn alter the
damping of the external force, e.g., low frequency sound waves in the
system. Provided these mechanical frequencies were small compared to
the rate of the control computations, the control force would be able
to track and respond to the external force at the desired frequency.

Another control task involves coupling the behavior of distinct parts
of the smart matter. For example, suppose we have two oscillators
whose behavior we want to have correlated. One way to do this is add a
control force $k (x_2-x_1)$ to the first oscillator, and $k (x_1-x_2)$
to the second. This gives an overall effective potential of

\begin{equation}
V_{\rm eff}(x_1,x_2) = \frac{1}{2} \Omega^2 (x_1^2 + x_2^2) - k x_1 x_2
\end{equation}

\noindent
with $\Omega = \sqrt{\omega^2 + k}$. By rotating the coordinate system
to use $y_1=(x_1+x_2)/\sqrt{2}$ and $y_2=(x_1-x_2)/\sqrt{2}$ this
becomes

\begin{equation}
V_{\rm eff}(y_1,y_2) = \frac{1}{2} \omega^2 y_1^2 + \frac{1}{2} (\omega^2 + 2k) y_2^2
\end{equation}

\noindent
Thus this coupling gives an additional restoring force acting on the
difference in position of the two oscillators, which will tend to keep
their positions correlated. Larger values of $k$ give stronger
correlations, but also require more control force.

In summary, these examples illustrate a variety of control methods
that can be applied. Although for simplicity we have considered
harmonic oscillations, the control computations could also allow
$k$ to vary with $x$, e.g., so the control is smaller when the system
is near the desired position. This amounts to adding aharmonicity to
the system. Even more generally, $k$ could also be time-dependent:
instead of the time-dependent force acting uniformly on the system as
treated above, this would allow more complex control forces. Although
more difficult to analyze, this flexibility greatly extends the
options for control strategies.

Therefore, one way to design quantum control methods is to apply
standard classical control algorithms, with some modification,
directly to quantum systems.  One major difference arises from the
fact that employing quantum computers requires the use of reversible
control programs, and as a result there is no possibility of creating
dissipative (non-conservative) control laws.  An, example, described
above is trying to control to the desired location of the center of
the packet by applying force to all $x$ values. Although conceptually
simple, it is by no means obvious that a control designed under the
assumption of zero packet width (i.e., a classical algorithm) will
continue to work for quantum systems. Moreover, some behaviors, such
as the width of the packet, have no classical analogs so to control
those aspects of the system will require uniquely quantum mechanical
control algorithms. Furthermore, control algorithms could also make
use of interference. An example would be combining several different
classical control algorithms so as to cancel out an undesired
behavior, even though each control by itself produces that
behavior. This gives additional options for control algorithms, beyond
attempting to use a single classical method. As another example, the
control could also manipulate the phase of the wave packet. A
superposition of such controls, e.g., based on different models of the
system, may be possible where the phase varies slowly near the correct
model, thus giving a strong contribution to the final result from the
control choice in the superposition that actually matches the system
parameters.

\section{Applications}

An important practical issue involves the construction of the devices
required for quantum smart matter.  While some progress has been
reported on the basic components of quantum
computers~\cite{barenco95,cirac95,sleator95}, quantum smart matter
also requires sensing and actuation abilities that can be coupled to
such computations. Thus progress in the development of quantum based
sensing and actuation is needed before applications, such as those
presented in this paper, become possible.  In addition, it is also
important to understand the relative time scales possible for quantum
computing and various physical behaviors that might be controlled.
This knowledge is required to develop effective sensing and actuation
mechanisms for any particular application.  Only when the computing is
relatively fast can we hope to construct smart matter for the behavior
in question.

To illustrate the concept of quantum smart matter, we present several
potential applications. These examples are based on applying
microscopic, decentralized or local, control to create a desired
macroscopic result. These applications are active camouflage; custom
manipulated material properties; control of certain chemical reaction
rates; and nonlinear, active springs for quantum machines.  The key
idea is to operate on the full quantum state through the control
program to produce a desired (global) behavior. To some extent,
classical controllers in smart matter could also be used. However, as
described above quantum computers allow for manipulating a wider range
of behaviors, and in particular to apply feedback without disrupting
the wave function.
 
\subsection{Active Camouflage}

Active camouflage takes advantage of the ability to create asymmetric,
optical potentials within materials~\cite{rosencher96}.  Manipulating
these potentials allows the controller to manipulate the optical
response to light striking that material and, thus, for example,
change the color with which it appears to an external observer. This
would allow quantum smart matter to determine the color of its
exterior surroundings, and modify the material potentials to reflect a
matching color. The end result is a material capable of actively
blending in with its surroundings like a chameleon.

There are several different time scales involved in this
example. First is the fast interaction of light with the
material. This is likely to be much faster than the time scale
governing computational speed. However, for this application, the
relevant physical time scale is the rate at which environment
conditions change, which is much slower. Hence substantial time could
be available for computation, perhaps using feedback to gradually
adjust the potentials to achieve a desired result.

\subsection{Active Materials}

Active materials employ quantum smart matter to manipulate their
lattice structure to customize their mechanical properties. This could
also be used to locally adjust the propagation of phonons or the
specific heat of the material, as well as to actively adjust the
material's mechanical behavior. These abilities would enable the
development of active thermal and acoustic isolators. In addition,
such behaviors could be employed to manipulate the displacement of a
structure under specific disturbance inputs, or to direct the thermal
stresses in a material to a desired location, similar to the way in
which a photocopier directs paper along a specific path.

Several applications could be accomplished using classical computers
for control.  For example, a signal might be sent to a portion of the
material to change the state of its control on the lattice structure,
modifying the stiffness, ductility, and strength of the material as a
result.  Such control inputs create a material that (at different
times) is both stiff and flexible, depending on what properties were
required, and where they were required.  One important application of
such a material is the forming of high strength alloys. In this
instance, a high strength alloy could actively be made more ductile
for forming, and then actively re-strengthened once in the proper
shape. Such an approach provides an elegant solution for the typical
difficulty encountered in industrially forming high strength materials
such as titanium.

Similar to the previous example, the relative time scales of physical
interaction and computational action are important. In this case, the
physical interaction occurs only as fast as the actuator bandwidth,
while the computational bandwidth is a function of the quantum
computers. The relevant time scale in this example is the desired
speed for adjusting the material properties. However, these
applications involve mechanical changes, which typically operate
slowly compared to compuational speeds.

\subsection{Adjusting Reaction Rates}

Chemical reaction rates can be affected by the local environment, such
as electric fields due to nearby ions. Smart matter capable of
adjusting fields could be used to modify reaction rates at a
surface. Alternatively, the material could be dispersed throughout
solutions containing the reactants. Coordinated programs running on
the individual pieces could then adjust reaction rates in a bulk
medium. This example illustrates that smart matter need not consist of
a single connected material since any necessary communication among
the controllers could use light or acoustic waves rather than wires. A
more subtle control would use local electric fields generated by the
active devices in a distributed version of controlling reactions
through external fields, a method that can also exploit quantum
interference~\cite{dahleh96,zhu95}. Chemical reactions can also be
controlled through mechanical forces~\cite{drexler92,gilman96}, thus
providing another path for smart matter to influence chemical
behaviors.  One possible application would be the control of certain
enzyme reactions. More generally, this would amount to a programmable
catalyst.

Since reaction rates are generally quite rapid, this application would
not involve active feedback response during reactions.  Rather the
control would take place by modifying the conditions for the reactions
at a slower time scale, leading to changes in the overall reaction
rate or the mix of products.  This could be done both through the
reduction of critical reaction elements and by using external field
changes previously determined to be useful. Feedback at this slower
time scale would still be useful for controlling complex reactions for
which accurate models are difficult to evaluate.

\subsection{Programmable Springs}

Controlled nonlinear springs could be created by using quantum smart
matter to control the bond strength between two (or several)
molecules. This control could also be used as a finer scale version of
current methods that modify molecular motion~\cite{dahleh96}.
Specifically, electric fields could be generated to control the bond
strength between two molecules, creating an active spring which might
be utilized in quantum machines.

The relevant time scale in this case is dependent on the application
of the active spring created.  Specifically, any application of such a
spring will have a maximum bandwidth that is necessary. As in the
active material example, this will involve mechanical time scales,
leaving plenty of time for computation.

\section{Conclusions}

We have described an application of quantum computers to control the
behavior of materials. Even fairly small computers, involving only a
few bits, may be able to produce useful new behaviors that would
otherwise be difficult to fabricate directly. The use of such
programmable materials could allow for experiments on the behavior of
many possible structures before deciding which few to actually attempt
to fabricate. In this way, quantum smart matter could be used as a
simulator for different quantum structures~\cite{lloyd96}. It could
also serve as an experimental platform for examining a range of
macroscopic quantum effects by introducing programmable correlations
in the overall quantum state of the material.

However, quantum computers face serious implementation difficulties of
decoherence and error control~\cite{landauer94a}, especially for
programs that require many steps. While substantial difficulities
remain before such devices can be constructed, there is encouraging
progress in the development of the basic components needed for quantum
computation~\cite{barenco95,cirac95,sleator95} and methods for error
control~\cite{shor95}. The simple individual programs useful for smart
matter may be less susceptible than others proposed for difficult
computational problems, but on the other hand any requirement for
communication over large distances will increase the difficulty of
avoiding undesired coupling to the environment. This provides another
reason to favor local, distributed control methods over the use of
global controls. Some local communications among the devices may
provide an application for proposals to transmit quantum
states~\cite{bennett93}.

We have suggested some possible applications of such capabilities, but
it remains to be seen whether the capacity to operate on
superpositions provides enough of an improvement over classical
computers to justify the difficulty in maintaining coherence. Finally,
if the promise of quantum computing is realized to improve
combinatorial search, e.g., for factoring~\cite{shor94,chuang95} or
more general cases~\cite{cerny93,hogg95d,grover96}, this capability
could also be used to give very complex potentials for the behavior of
matter.

\section*{Acknowledgements}

We thank B. Huberman and R. Merkle for helpful comments on this work.

\end{document}